\documentclass[pre,floats,preprint,superscriptaddress,showpacs,usenames]{revtex4}
\usepackage{amsmath}
\usepackage{graphicx}

\usepackage[dvipsnames]{color}

\newcommand{\be}{\begin{equation}} 
\newcommand{\ee}{\end{equation}}
\newcommand{\bea}{\begin{eqnarray}}   
\newcommand{\eea}{\end{eqnarray}}

\newcommand{\rr}{{\bf r}}
\newcommand{\NN}{{\bf \nabla}}
\newcommand{\FF}{{\bf F}}
\newcommand{\GG}{{\bf G}}

\newcommand{\vv}{{\bf v}}
\newcommand{\vva}{{\bf v}^{\alpha}}

\newcommand{\fa}{f^{\alpha}}
\newcommand{\na}{n^{\alpha}}
\newcommand{\nb}{n^{\beta}}
\newcommand{\rhoa}{\rho^{\alpha}}

\newcommand{\uu}{{\bf u}}
\newcommand{\uua}{{\bf u}^{\alpha}}

\newcommand{\uai}{u^{\alpha} _i}
\newcommand{\uaj}{u^{\alpha }_j}

\newcommand{\uub}{{\bf u}^{\beta}}
\newcommand{\ma} {m^{\alpha}}

\newcommand{\mb} {m^{\beta}}

\newcommand{\sab}{\sigma_{\alpha\beta}}
\newcommand{\gab}{g_{\alpha\beta}}
\newcommand{\muab}{\mu_{\alpha\beta}}

\newcommand{\qq} {{\bf q}}
\newcommand{\bk}{\hat{\bf s}}

\newcommand{\ww}{{\bf w}}

\begin{document}
\date{\today}
\title{Multicomponent Diffusion in Nanosystems}

\author{Umberto Marini Bettolo Marconi\footnote[3]
{(umberto.marinibettolo@unicam.it)}
}

\address{ Scuola di Scienze e Tecnologie, 
Universit\`a di Camerino, Via Madonna delle Carceri, 62032 ,
Camerino, INFN Perugia, Italy}

\author{Simone Melchionna}
\address{CNR-IPCF, Consiglio Nazionale delle Ricerche, Italy}
\begin{abstract}
We present the detailed analysis of the diffusive transport of spatially 
inhomogeneous fluid mixtures
and the interplay between structural and dynamical properties varying on the atomic scale.
The present treatment is based on different
areas of liquid state theory, namely kinetic and density functional 
theory and their implementation as an effective numerical method via the
Lattice Boltzmann approach.
By combining the first two methods it is possible to
obtain a closed set
of kinetic equations for the
singlet phase space distribution functions of each species.
The  interactions among particles are 
considered within a self-consistent approximation 
and the resulting effective molecular fields are analyzed.
We focus on  multispecies diffusion in systems with
short-range hard-core repulsion between particles of unequal sizes and
weak attractive long-range interactions.
As a result,
the attractive part of the potential
does not contribute explicitly to viscosity  but  to diffusivity and 
the thermodynamic properties.
 Finally, we obtain a practical scheme to solve the kinetic equations by employing a discretization procedure derived from the Lattice
 Boltzmann approach.  Within this framework, we present numerical data  concerning
the mutual  diffusion properties both in the case of a quiescent bulk fluid  and shear flow 
inducing Taylor dispersion.

 \end{abstract}

\maketitle
  
%        SECONDO ARTICOLO

\section{Introduction}

Modern applications in science, medicine
and technology require a better understanding 
of the molecular mechanisms
controlling the flow of liquids near solid substrates and at interfaces \cite{sparreboom,bocquet,schoch,Bruus}.
 It is well known that structural and transport properties of highly 
confined fluids or near free surfaces differ from their bulk behavior due to the 
large surface to volume ratio \cite{rauscher}.
Many phenomena occurring at molecular scales such as diffusion, mixing,
shear thinning and lane formation
involve the interplay between 
microscopic structural and transport properties,
which need the investigation
of the long-time flow behavior. This task is computationally very demanding for
approaches such as Molecular Dynamics,
so that alternative methods are desirable. Some of these alternative approaches  are intermediate between macroscopic thermodynamic
and truly microscopic methods and have
the scope to incorporate molecular details, at the price of 
a limited amount of numerical effort. 
Among these, the dynamic density functional theory 
(DDFT) and Direct Simulation Monte Carlo (DSMC) are prominent numerical methods.
DSMC is a direct particle simulation method based on kinetic theory and its basic
idea is to follow the trajectories of a large number of statistically representative particles
and stochastical collisions are modeled  using scattering probabilities. 
It gives results which are accurate on scales shorter than the mean free path \cite{Garcia,Santos}.
The DDFT assumes that the evolution of the system is determined by
a ``thermodynamic force'', which is
 the functional derivative of the free energy functional ${\cal F}$
with respect the local  density
 \cite{ArcherEvans,Tarazona,Lowen,rauscher2,wu}. In DDFT the state of the solute particles at time $t$ is described 
by the average density $n(\rr,t)$ while  the solvent is assimilated to  a continuum whose
interactions  with the solute are modeled via a stochastic heat-bath mechanism. However, this approach is inappropriate
to describe the hydrodynamic behavior of simple liquids and liquid mixtures 
since within the DDFT picture the momentum transport
can only occur via diffusion, but not via convection \cite{umbmnogalilei,Espanol,Archer2009,Lausanne2010}.

On the contrary,  kinetic methods extending the
Boltzmann equation to the dense fluid regime,
can in principle describe both the thermodynamic and the hydrodynamic behavior of simple fluids.
In spite of its great historical relevance in statistical physics, the Boltzmann-Enskog
approach has rarely enjoyed the due attention in the area of inhomogeneous fluids,
apart from some notable exceptions \cite{Zhaoli,Greci}. The reason perhaps being
that, under spatially  inhomogeneous conditions, numerical solutions of the equation are impractical.
However, the situation has changed with the advent of modern lattice techniques for
solving the Boltzmann equation, collectively named the Lattice Boltzmann Method (LBM) \cite{LBgeneral,shanchen,Doolen,Luo}.
The simultaneous discretization of positional and translational degrees of freedom
enables the efficient solution of such an equation by brute force. On the other hand,
the application of the LBM to small systems
is usually considered to be  outside the realm of applicability
of kinetic methods, but routinely treated within the DDFT approach, provided that 
the considered systems are not too far from local equilibrium conditions.

In a series of recent papers we proposed a formulation of the Boltzmann-Enskog theory
which is thermodynamically consistent, gives satisfactory values of the transport coefficients,
and lends itself to numerical solutions within the LBM framework \cite{Melchionna2008,Melchionna2009,JCP2011}.
The method proved to provide reliable results in simple geometrical set-ups and
was later extended  to  multicomponent fluids and to their rich and fascinating phenomenology.
In the present paper, we  investigate further issues
related to the multicomponent system with special attention 
to the diffusive behavior.

Following few significant studies published on the subject
\cite{Davis1,Davis2,Pozhar,nicholson,Rice}, but differing from ours in the 
treatment of the short range correlations,  
we represent the evolution of the system in terms of the
singlet phase space distribution functions, $\fa(\rr,\vv,t)$, referring to species $\alpha$.
% The method is specially suited for studying  transport phenomena
% of fluids with underlying Newtonian dynamics.
The governing kinetic equations and the  balance equations 
for the density and momentum current of the individual species, obtained in previous work
employing the multicomponent extension of the method of Dufty and coworkers \cite{Brey},
are briefly summarized in Sec. \ref{Theory} to render the paper self-contained. These balance equations involve 
different kinds of forces  which are the subject of the analysis of Sec. \ref{forceanalysis} resulting in an
identification of hydrostatic, capillary, viscous and drag forces in terms of microscopic parameters.
In Sec. \ref{binarymixture} we specialize the theory to a binary mixture
and in Sec. \ref{homogeneousdiffusion} we turn our attention to  the evolution of the  local concentration and
show how to derive microscopically the advection-diffusion equation under suitable assumptions.
In Sec. \ref{hydrodynamicanalysis} we perform an hydrodynamic analysis of the coupled
set of balance equations in order to illustrate the response of a nearly homogeneous mixture to small
deviations from the local equilibrium state.
Finally in Sec. \ref{Numerical} we solve numerically the transport equation utilizing the  extension of  the Lattice Boltzmann (LBM)
method  proposed in \cite{JCP2011},  where the positions  are discretized on a lattice
and the velocities  discretized over a small basis set. This strategy renders the  computations efficient 
and numerically stable.
The method was validated against the diffusion of a small periodic inhomogeneity
for several values of the bulk parameters.
We have also numerically studied  the
coupling between microscopic diffusion and a non uniform velocity field,
a problem  known as Taylor dispersion \cite{taylor}.
A numerical comparison between the analytical work and the numerical solution 
of the model shows a satisfactory agreement with the theoretical predictions.
We conclude this section by discussing the role of the attractive tails
in determining the diffusion coefficient.
Finally, Sec.\ref{Conclusion} contains some concluding remarks.

%The present work deals with mixtures,  important to a great variety of natural and technological systems,
%and represent an extension of  our recent treatment 
%of one component fluids. With respect to ref. \cite{precedente} 
%incorporates temperature fluctuations and a 
%detailed treatment of  mutual and thermal diffusion, two important
%phenomena which have no counterpart in pure fluids.

%%%%%%%%%   THEORY

\section{Multicomponent Transport equation}
\label{Theory}

In the present paper, we shall employ a recent method to describe the isothermal 
transport properties of a mixture
\cite{JCP2011}.
The idea is 
to simplify the transport problem by approximating the interaction term
in such a way that non-local correlations, giving rise to the 
microscopic structure of the fluid, are taken into account.  The
approximation determines a non trivial dependence of the transport coefficients on the density profiles.
In a recent paper \cite{JCP2011} we have derived the evolution of the singlet phase-space distribution function,
$\fa(\rr,\vv,t)$, characterizing the state of species $\alpha$, of mass $\ma$, in a M-component fluid mixture, 
by means of the following transport equation:
\bea
&&
\frac{\partial}{\partial t}\fa(\rr,\vv,t) +\vv\cdot\NN \fa(\rr,\vv,t)
+\frac{\FF^{\alpha}(\rr)}{\ma}\cdot
\frac{\partial}{\partial \vv} \fa(\rr,\vv,t)=\nonumber\\
&&
-\omega[ \fa(\rr,\vv,t)- \psi^{\alpha}_{\perp}(\rr,\vv,t)]+\frac{{\bf \Phi}^{\alpha}(\rr,t) }{k_B T} \cdot(\vv-\uu(\rr,t)) 
 \psi^{\alpha}(\rr,\vv,t) ,
\nonumber\\
\label{evolution}
\eea
where $\FF^\alpha$ is an external body force acting on species $\alpha$, $T$  the uniform temperature of the
system and $k_B$ the Boltzmann constant.
The central quantity of eq. (\ref{evolution}) is ${\bf \Phi}^\alpha(\rr,t)$, which
bears the result of collisions between particles, and whose details will be given below.
In addition, $\psi^\alpha$ is the local Maxwellian equilibrium of specie $\alpha$,
\be
\psi^{\alpha}(\rr,\vv,t)=\na(\rr,t)[\frac{\ma}{2\pi k_B T}]^{3/2}\exp
\Bigl(-\frac{\ma(\vv-\uu(\rr,t))^2}{2 k_B T} \Bigl)
\label{psia}
\ee
and the distribution $\psi_\perp^\alpha$
shares to the same average density and velocity as the actual distribution $\fa$:
\bea
&&
\psi^{\alpha}_{\perp}(\rr,\vv,t)=\psi^{\alpha}(\rr,\vv,t) \Bigl\{1+
\frac{\ma(\uua(\rr,t)-\uu(\rr,t))\cdot(\vv-\uu(\rr,t))}{k_B T}\Bigl\} \nonumber\\
\label{prefactor}
\eea
%%%%%%%%%%%%%%%%%%

Eqs. \eqref{evolution}-\eqref{psia} contain  the fields $\na,\uu^\alpha,\uu$, the average partial number density
of the component $\alpha$, its average velocity and the barycentric velocity of the mixture, respectively.
The first two quantities are defined by:
\be
\left( \begin{array}{cc}
n^\alpha(\rr,t) \\
n^\alpha(\rr,t) {\bf u}^{\alpha}(\rr,t)\\
 \end{array} \right) =
\int d\vv
\left( \begin{array}{cc}
1   \\
\vv  \\
 \end{array} \right)  
f^{\alpha}(\rr,\vv,t).
\label{colonna}
\ee
One also needs to specify  the partial mass density,
$
\rhoa(\rr,t)=\ma \na(\rr,t),
$
the global number density, $n(\rr,t)=\sum_\alpha \na(\rr,t)$, the global mass density 
\be
\rho(\rr,t)=\sum_\alpha \rhoa(\rr,t)
\label{massdensity}
\ee
and the barycentric average velocity at position $\rr$:
\be
\uu(\rr,t)=\frac{\sum_{\alpha}\rhoa(\rr,t)\uua(\rr,t)}
{\sum_{\alpha}\rhoa(\rr,t)}
\ee

Eq. \eqref{evolution} is an approximate isothermal representation of the revised Enskog theory (RET) kinetic equation
\cite{vanbeijeren} where, in order to obtain a workable scheme, the non-linear collision operator
has been replaced by the two terms featuring in the r.h.s. of the equation.
It is a simplified representation of the multicomponent RET for hard sphere mixtures, which
contains two features  that go beyond the standard Boltzmann equation approach \cite{Lopezdeharo}. 
The colliding particles 
are separated by a distance equal to the sum of their radii and the collision frequency is modified to take into account the
excluded volume effect through the introduction of the pair correlation function at contact in the collision integral. 
Such a pair correlation function depends on the densities through a smoothing procedure. 
The first term in the l.h.s. of  eq. \eqref{evolution} describes the fast relaxation process towards local equilibrium and
represents in an approximate fashion the non-hydrodynamic part of the  collision operator.
It contains $\omega$, a collision  frequency assumed to be the same for all species.

The form of the first term in the r.h.s of eq \eqref{evolution} is clearly reminiscent of the  Bhatnagar-Gross-Krook  (BGK) relaxation 
term employed in the treatment of one-component systems \cite{BGK}. It contains an additional 
factor making the difference between $\psi^\alpha_\perp$ and $\psi^\alpha$. 
The factor multiplying the Maxwellian in eq. (\ref{prefactor}) 
serves to "orthogonalize" the term $-\omega[ \fa- \psi^{\alpha}_{\perp}]$  
to the term containing the effective fields, ${\bf \Phi}^\alpha$, as specified below.
Such a modification
is necessary in order to produce the correct balance equation for the partial momentum
and to obtain the correct form of the momentum equation for the individual components
(see eq. \eqref{momentcomponent}).

%%%%%%%%%%%%%%%%%%%   BALANCE EQUATIONS

%\subsection{Balance equations for individual species}
In the following, we consider the evolution of the partial density
and of partial momentum current. The first is
obtained by integrating eq. \eqref{evolution} w.r.t. the velocity 
\be
\frac{\partial}{\partial t}\rhoa(\rr,t) +\nabla\cdot \Bigl(\rhoa(\rr,t) \uu(\rr,t)\Bigl)+
\nabla\cdot \Bigl(\rhoa(\rr,t)(\uua(\rr,t)- \uu(\rr,t)\Bigl)=0 ,
\label{continuity}
\ee
where the last term in eq. (\ref{continuity}) 
is the so-called dissipative diffusion current,
measuring the drift of the $\alpha$-component with respect to the center of  mass velocity.

Multiplication of eq. \eqref{evolution} by $\ma \vv$ and integration  w.r.t. $\vva$ yields
the balance equation for the momentum of the species $\alpha$:
\bea
&&
\frac{\partial}{\partial t}[\rhoa(\rr,t)\uaj(\rr,t)]+ 
\nabla_i \Bigl(\rhoa(\rr,t) \uai(\rr,t) \uaj(\rr,t)
-  \rhoa(\rr,t)(u^{\alpha}_i(\rr,t)-u_i(\rr,t))( u^{\alpha}_j(\rr,t)-u_j(\rr,t))\Bigl)=
\nonumber\\
&& 
-\nabla_i \pi_{ij}^{\alpha}(\rr,t)+ \frac{F^{\alpha}_j(\rr)}
{\ma}\rhoa(\rr,t)+
 \frac{ \Phi^{\alpha}_{j}(\rr,t)}{\ma}\rhoa(\rr,t) ,
\label{momentcomponent}
\eea
where
\be
\pi_{ij}^{\alpha}(\rr,t)=\ma\int d\vv (v_i-u_i)(v_j-u_j)\fa(\rr,\vv,t)
\label{pressurekin}
\ee
represents the kinetic contribution of component $\alpha$ to the pressure tensor. Here and in the
following the Einstein convention on repeated indices is employed.

%%%%%%%%%%%%% FORCE ANALYSIS

\section{Force analysis}
\label{forceanalysis}

In ref. \cite{JCP2011} we derived an explicit expression for the effective fields,
  ${\bf \Phi}^{\alpha}(\rr,t)$, for a model with repulsive hard sphere potentials of different diameters, $\sigma_{\alpha\alpha}$
 and masses $\ma$, plus
 long range attractive interactions with associated potential term $U^{\alpha\beta}$.
The central notion is that this quantity is a functional of the density and velocity 
of each species. 
By treating the repulsive contribution in the framework of  the revised Enskog theory  \cite{vanbeijeren},
and the attractive term within the random phase approximation (RPA) \cite{hansen},
the effective field is represented
as a sum of forces of different nature:
 \be
{\bf \Phi}^{\alpha}(\rr,t)= \FF^{\alpha,mf}(\rr,t)+\FF^{\alpha,drag}(\rr,t)+\FF^{\alpha,visc}(\rr,t) .
\label{splitforce}
\ee
The first term represents the force acting on species $\alpha$ at position $\rr$ due to
the influence of all remaining particles in the system, and is the gradient of the so-called potential of mean force.
When the system is in thermodynamic equilibrium such a force is related to the excess of the chemical
potential \cite{Evans1,Carey} over its ideal gas value,  $ \mu_{exc}^\alpha$, of the $\alpha$ component by the relation:
%% POTENTIAL OF MEAN FORCE
\be
\FF^{\alpha,mf}(\rr,t)= -\NN \mu_{exc}^{\alpha}(\rr,t).
\label{potchimico}
\ee
 Explicitly, using the form of the RET collision term and an attractive potential tail, we obtain 
the following representation 
\be
\FF^{\alpha,mf}(\rr,t)=-k_B T\sum_\beta\sab^2 
\int d\bk \bk
g_{\alpha\beta}(\rr,\rr+\sab\bk,t)
n_{\beta}(\rr+\sab\bk,t)+\sum_\beta \GG^{\alpha\beta}(\rr,t)
\label{potmeanforce}
\ee
where $\sab=(\sigma_{\alpha\alpha} +\sigma_{\beta\beta})/2$
and the integration in the first term of the r.h.s. is over the unit spherical surface, while the last term represents  the molecular fields associated with the attractive forces:
\be
\GG^{\alpha\beta}(\rr,t)=
- \int dr' \nb(\rr',t)\gab(\rr,\rr')\NN_r U^{\alpha\beta}(\rr-\rr')
\ee
%%%%%%%%%%%%%%%%%%%%%%%%%%%%%%%%%%CCCCCCCCCCCCCCCCCCCCC

% DRAG FORCE

The second and third terms of eq. \eqref{splitforce} carry a functional dependence on the 
velocities, contributions that are neglected in semi-macroscopic models of
single or multicomponents \cite{LeeFischer}. These terms are crucial for the correct
characterization of dissipation and diffusion in the condensed state and result in
density-dependent transport coefficients.

The second term in the r.h.s. of  eq. \eqref{splitforce} is the  drag force exerted by unlike species on the particle $\alpha$
in reason of their different drift velocities:

\be
\FF^{\alpha,drag}(\rr,t)= 
-\sum_\beta{\bf \gamma}^{\alpha\beta} (\rr,t)  (\uua(\rr,t)-\uub(\rr,t)),
\label{gammaab}
\ee
where we have introduced an inhomogeneous friction tensor via the equation:
\be
\gamma_{ij}^{\alpha\beta}(\rr,t)=2\sigma_{\alpha\beta}^2 \sqrt{\frac{2\muab k_B T}{\pi} }
\int d\bk s_i s_j
\gab(\rr,\rr+\sigma_{\alpha\beta}\bk,t)
\nb(\rr+\sigma_{\alpha\beta}\bk,t).
\label{tensorgamma}
\ee
Finally, the last  term in the r.h.s. of  eq. \eqref{splitforce}  represents the viscous force due to the presence of velocity gradients:
% VISCOUS FORCE
\be
\FF^{\alpha,visc}(\rr,t)=
\sum_\beta 2\sab^2 \sqrt{\frac{2\muab k_B T}{\pi} }
\int d\bk \bk
g_{\alpha\beta}(\rr,\rr+\sab\bk,t)
\nb(\rr+\sab\bk,t)
 \bk\cdot
(\uub(\rr+\sab\bk)-\uub(\rr)) ,
\label{viscousforce}
\ee
where $\gab(\rr,\rr+\sab\bk,t)$ is the pair correlation functions at contact ($|\rr-\rr'|=\sab$) and $\muab$ 
is the reduced mass $\muab=\frac{\ma \mb}{\ma+\mb}$ for the colliding pair.

%%%%%%%%%%%%%%%%%%%%%%%%%%%%%
% \subsection{Linear estimate of the force terms for weak perturbations}

In the case of weak spatially periodic deviations from the homogeneous reference state,
it is possible to derive explicit expressions for the effective forces discussed above.
At first, let us consider a slowly varying periodic variation of the densities of the two species of the form:
\be
\na(\rr,t)= \na_0(t) +\delta \na(t) e^{i {\bf q \cdot r}}
\ee
where $q \sab<<1$ and $\na_0$ are uniform densities. By substituting such a density profile into
eq. \eqref{potmeanforce}  and expanding the resulting integrals up to second order in the parameter $q\sab$, we find the expression:
%\be
%\FF^{\alpha,mf}(\rr,t) \simeq -k_B T e^{i {\bf q \cdot } \rr}\sum_\beta\sab^2 [g^{bulk}_{\alpha\beta}+\frac{\partial g^{bulk}_{\alpha\beta}}{\partial \nb}]
%\delta \nb(t) \int d\bk \bk e^{i {\sab\bf q \cdot }\bk}
%-i \qq  e^{i {\bf q \cdot } \rr} \sum_\beta \delta \nb(t)\int d\rr'U^{\alpha\beta}(\rr')e^{i {\bf q \cdot } \rr'}
%\ee
%or
\be
\FF^{\alpha,mf}(\rr,t) \simeq -i \qq e^{i {\bf q \cdot } \rr}  \sum_\beta  \delta \nb(t)
\Bigl [k_B T \frac{4 \pi}{3}
\Bigl(\sab^3 g^{bulk}_{\alpha\beta}+
\frac{1}{2}\sum_\gamma n^\gamma\sigma_{\alpha\gamma}^3\frac{\partial g^{bulk}_{\alpha\gamma}}{\partial \nb}\Bigl) -  
w_0^{\alpha\beta} +\frac{1}{2}w_2^{\alpha\beta} q^2  \Bigl]
\label{diciotto}
\ee
with $ w_n^{\alpha\beta}=\int d\rr |\rr|^n U^{\alpha\beta}(\rr)$.
The last  term in   the l.h.s. of eq. \eqref{diciotto} corresponds to 
the contribution to the local force acting on the species $\alpha$ stemming from the
attractive interactions \cite{Carey}.

Similarly,  we estimate the viscous force by considering uniform densities and a weak periodic velocity field
$\uu(\rr,t)= \uu(t) e^{i {\bf q \cdot r}}$, with $\uu^A=\uu^B$. We find
%\be
%\FF^{\alpha,visc}(\rr,t)=
%2   \sum_\beta \sab^2 \nb_0 \gab\sqrt{\frac{2\mu_{AB} k_B T}{\pi} }\uu(t) e^{i {\bf q \cdot r}}
%\int d\bk \bk \bk\cdot
%(e^{i {\sab\bf q \cdot }\bk}-1) 
%\label{viscousforce}
%\ee
\be
\FF^{\alpha,visc}_{\perp}(\rr,t)\simeq
-\frac{4\pi}{15} q^2
\uu_{\perp}(t) e^{i {\bf q \cdot r}} \sum_\beta \sab^4 \nb_0 \gab\sqrt{\frac{2\mu_{AB} k_B T}{\pi} },
\label{viscousforce1}
\ee
and
\be
\FF^{\alpha,visc}_{||}(\rr,t)\simeq
-\frac{4\pi}{5} q^2
\uu_{||}(t) e^{i {\bf q \cdot r}} \sum_\beta \sab^4 \nb_0 \gab\sqrt{\frac{2\mu_{AB} k_B T}{\pi} }
\label{viscousforce2},
\ee
where we have considered the parallel and the perpendicular part of the velocity with respect to the
wave-vector $\qq$.
%Introduce $\eta^C$.
%For small deviations from the uniform state
%$(\rho(\rr,t)=\rho_0,\uu(\rr,t)=0,\uu^\alpha(\rr,t)=0,\rhoa(\rr,t)=c_0 \rho_0)$ as
%shown in previous articles \cite{Melchionna2Finally the drag force can be estimated by imposing constant densities and different velocities for the two species
%\be
%\FF^{\alpha,drag}(\rr,t)=  -(\uu^A(\rr,t)-\uu^B(\rr,t)) \gamma^{AB} 
%\label{gammaab}
%\ee
As a result we obtain
\be
\gamma^{\alpha\beta}=\frac{8}{3} \sqrt{2 \pi\mu_{\alpha\beta} k_B T}
g_{\alpha\beta}n^\beta\sigma_{\alpha\beta}^2 .
\ee

%CCCCCCCCCCCCCCCCCCCC%

\section{The binary mixture}

\label{binarymixture}
In order to proceed with analytical work 
it is more convenient to use  as variables the local mass density and local  momentum variables
together with concentration variables. The new equations  can be obtained by combining appropriately 
equations \eqref{continuity} and \eqref{momentcomponent}.

%%%%%%%%%%%%%%%%%%%%%%%%%%%%%%%%%%

By specializing to a binary mixture, AB, the local concentration is defined as
\be
c(\rr,t)=\frac{\rho^A(\rr,t)}{\rho(\rr,t)}
\ee
%and the local mass density, $\rho(\rr,t)$.
From the evolution equations \eqref{continuity}  for the partial densities,
the mass continuity equation reads
% CONTINUITY MASS
\be
\partial_{t}\rho(\rr,t)+\nabla\cdot \Bigl( \rho(\rr,t) \uu(\rr,t)\Bigl)=0.
\label{densityconservation}
\ee 
and the conservation law for the local concentration
% CONTINUITY CONCENTRATION
\be
\frac{\partial}{\partial t} c(\rr,t) +\uu(\rr,t) \cdot \nabla c(\rr,t)+\frac{1}{\rho}
\nabla\cdot \Bigl(\rho(\rr,t)c(\rr,t)(1-c(\rr,t)){\bf w}(\rr,t)\Bigl)=0 ,
\label{continuityconcentration}
\ee
where we have introduced the velocity difference 
\be
 {\bf w}(\rr,t)\equiv \uu^A(\rr,t)-\uu^B(\rr,t).
 \ee

%C MOMENTO
Using eq.\eqref{momentcomponent}, the equation expressing the total momentum balance reads
\bea
&&\partial_{t}u_j(\rr,t)+ u_i(\rr,t)\nabla_i u_j(\rr,t)+\frac{1}{\rho}\nabla_i \pi^{(K)}_{ij}\nonumber\\
&& -\frac{1}{\rho}\Bigl(n^A(\rr,t)[F^{A }_j(\rr) +F^{A,mf}_{j}(\rr,t) +F^{A,visc}_{j}(\rr,t) ]+
 n^B(\rr,t)  [F^{B }_j(\rr) + F^{B,mf}_{j}(\rr,t) +F^{B,visc}_{j}(\rr,t)]  \Bigl)=0. \nonumber\\ 
 \label{globalmomentumcont}
\eea
To proceed further,
it is convenient to define the total local chemical potential of each species $A(B)$ through the equation:
\be
\nabla_j \mu^{A(B)}(\rr,t)\equiv \frac{1}{n^{A(B)}(\rr,t)}\nabla_i \pi_{ij}^{A(B)}(\rr,t)\delta_{ij}- F^{A(B),mf}_{j}(\rr,t) ,
\label{chema}
\ee
where  
%to write eq. \eqref{chema} 
we used eq. \eqref{potchimico} for the non ideal part and the relation between the ideal gas pressure and the
chemical potential of an ideal gas.
In the isothermal system, the gradient of the total thermodynamic pressure is defined as
\be
\nabla_j P(\rr,t) \equiv n^A(\rr,t)\nabla_j \mu^A(\rr,t)+n^B(\rr,t)\nabla_j \mu^B(\rr,t)
\label{gibbsduhem}
\ee
that can be seen as a special case of the Gibbs-Duhem equation.

%%%%%%%%%%
In the following we shall use the fact that
the kinetic contribution to the gradient of the pressure tensor, $\pi^{(K)}_{ij}= \pi^{A}_{ij} +\pi^{B}_{ij} $, can be written as:
\be
\nabla_i \pi^{(K)}_{ij} \simeq \delta_{ij}\nabla_j P_{id}
 -\eta^{(K)}\Bigl(\frac{1}{3}\nabla_i\nabla_j u_i +\nabla_i^2 u_j\Bigl) ,
\ee
with $P_{id}=k_B T(n^A(\rr,t)+n^B(\rr,t))$
and $\eta^{(K)}=\frac{k_B T}{\omega}(n^A(\rr,t)+n^B(\rr,t))$.
As shown in Ref. \cite{molphys2011}, in the limit of small gradients also 
the non-ideal contribution to the  viscous force in  the momentum equation can be written as:
\be
\sum_\alpha\na(\rr,t)\FF^{\alpha,visc}(\rr,t)    
\simeq-\eta^{(C)}\nabla^2\uu-(\frac{1}{3}\eta^{(C)}+\eta_b^{(C)})\nabla(\nabla\cdot\uu)
\ee
where the non-ideal contribution to the shear viscosity is 
\be
\eta^{(C)}=\frac{4}{15}\sum_{\alpha\beta}\sqrt{2\pi\muab k_B T}\sab^4 \gab \na_0\nb_0 ,
\ee
while the bulk viscosity is
\be
\eta_b^{(C)}=\frac{5}{3}\eta^{(C)} .
\ee
Notice that, within our approximations
the kinetic contribution to the  bulk viscosity vanishes:
$\eta_b^{(K)}=0$.

%% FOURTH EQUATION
In order to
derive an expression for ${\bf w}$, we compute the difference between the
velocities of the two components using eq. (\ref{momentcomponent}), 
and derive the following equation :
\bea
&&
\frac{\partial}{ \partial t}w_j(\rr,t)+ \Big[
u^M_i(\rr,t) \nabla_i w_j(\rr,t)+w_i(\rr,t) \nabla_i  u^M_j(\rr,t)\Bigl)
\nonumber\\&&
- \frac{1}{ \rho^A}
\nabla_i \Bigl( \frac{\rho^A (\rho^B)^2}{ \rho^2} 
w_i(\rr,t) w_j(\rr,t)\Bigl)+
\frac{1}{ \rho^B}
\nabla_i \Bigl( \frac{\rho^B (\rho^A)^2}{ \rho^2} 
w_i(\rr,t) w_j(\rr,t)\Bigl)\Bigl]
=
\nonumber\\
&& 
-\Bigl(\frac{1}{\rho^A}\nabla_i \pi_{ij}^A
-\frac{1}{\rho^B} \nabla_i \pi_{ij}^B \Bigl)
+\Bigl( \frac{\Phi^A_{j}(\rr,t)+F^{A}_j(\rr)}
{m^A}- \frac{\Phi^B_{j}(\rr,t)+F^{B}_j(\rr)}{m^B}\Bigl),
\nonumber\\
\label{diffmoment}
\eea
with the abbreviation  $\uu^M\equiv (\uu^A+\uu^B)/2$ .
Before studying such an equation we shall make some considerations.

%%%%%%%%%%%%%%%%%%%%%%%%%%%%%%%%%%%%%%%
%

\subsection{Homogeneous diffusion}
\label{homogeneousdiffusion}
The phenomenon of   
multicomponent diffusion has ever since attracted a vivid interest
\cite{Chapman,Degroot,Ferziger,Pina,Tham,Gross,Garzo,Andries}.
We shall specialize the discussion to the binary mixture 
and consider first a system where diffusion is the dominant mechanism to restore 
equilibrium, assuming that the global velocity of the fluid is nearly uniform. 

%Later we  shall discuss the diffusion under inhomogeneous conditions. 
%The presence of two different species entails the diffusion coefficient turns out to be the one predicted by the Enskog theory \cite{Chapman}.

In eq. \eqref{momentcomponent} the inertial term, $\nabla_i[\rho^\alpha u^\alpha_i u^\alpha_j]$,
 is small with respect to the terms associated with the viscous component of the kinetic and 
 potential parts of the pressure tensor (see eq. \eqref{viscousforce} ),
 $(\nabla_i \pi_{ij}^{\alpha}+\na F^{\alpha,visc}_j)$,  and their ratio is  given by:
\be
\frac{\nabla_i[\rhoa u^\alpha_i u^\alpha_j] }{(\nabla_i \pi_{ij}^{\alpha}+\na F^{\alpha,visc}_j)}\simeq \frac{\rho u^2/L}{\eta u/L^2}=\frac{\rho u L}{\eta }={\cal {R} },
\ee
where $L$ is the typical spatial scale of the gradients, $u$ the  velocity of the flow and $\eta$ the shear viscosity  and ${\cal R}$
is the Reynolds number. 
%Since we need a closed expression for the diffusive current we must solve eq. \eqref{diffmoment} 
In  eq. \eqref{diffmoment} the viscous term is also negligible with respect to the
chemical potential term :
\be
\frac{\eta \nabla^2 u}{n\nabla \mu} \simeq\frac{\eta u/L^2}{\rho c_s^2 L}=\frac{1}{\cal {R} }\frac{u^2}{c_s^2}
=\frac{Ma^2}{\cal R}
\label{Mare}
\ee
where $c_s$ is the sound velocity and $Ma=u/c_s$ the Mach number.
Thus, in the regime of low velocities the ratio
\eqref{Mare} is small and the viscous force in   eq. \eqref{diffmoment} can be safely neglected.
Using eq. \eqref{gammaab}, we rewrite eq. \eqref{diffmoment} as
%%%%%%%%%%%%%%%%%%%%%%
\bea
\frac{\partial}{ \partial t}w_i(\rr,t)+\nabla_i\mu_D(\rr,t)+\Bigl(
\frac{1}{m^A}\gamma_{ij}^{AB}(\rr,t) +\frac{1}{m^B}\gamma_{ij}^{BA}(\rr,t)\Bigl)
  w_j(\rr,t)=\Bigl( \frac{F^{A}_i(\rr)}{m^A}- \frac{F^{B}_i(\rr)}{m^B}\Bigl)
 \label{wequation}
\eea
where the appropriate thermodynamic field, $\mu_D$,  conjugated to
the concentration variable, $c=\rho^A/\rho$, is the difference in the chemical potentials per unit mass of the two components \cite{hansen}  defined  
as:
\be
\nabla_j\mu_D(\rr,t)\equiv \frac{1}{m^A}\nabla_j \mu^A(\rr,t)
-\frac{1}{m^B} \nabla_j \mu^B(\rr,t) .
\label{mudif}
\ee

In the homogeneous case the friction tensor is isotropic and diagonal
and can be written as
\be
\gamma\equiv \frac{1}{m^A}\gamma_{ii}^{AB}(\rr,t) +\frac{1}{m^B}\gamma_{ii}^{BA}(\rr,t)
=\frac{8}{3}\rho\frac{\sqrt{2\pi\mu_{AB} k_B T }}{m^A m^B} g^{bulk}_{AB} \sigma_{AB}^2 .
\label{gammasymmetrized1}
\ee

It can also be assumed that in eq. \eqref{wequation} the variation in time of $\ww$ is slow, so that
\be
\ww(\rr,t)=-\frac{1}{\gamma}\Bigl\{ \nabla\mu_D(\rr,t)-\Bigl( \frac{\FF^{A}(\rr)}{m^A}- \frac{\FF^{B}(\rr)}{m^B}\Bigl)\Bigl\}
\label{wapprox}
\ee
Using the Gibbs-Duhem equation, eq.\eqref{gibbsduhem}, the chemical potential difference can be expressed as
\be
\nabla \mu_D=\frac{\rho}{n m^A m^B}\Bigl(\nabla(\mu^A-\mu^B)+(m^B-m^A)\frac{1}{\rho}\nabla P\Bigl)
\ee
by substituting into eq. \eqref{wequation}, we find that in stationary conditions, 
\be
\ww(\rr,t)=-\frac{D^{AB}}{k_B T}
\Bigl\{ \NN( \mu^A- \mu^B)+ (m^B-m^A) \frac{1}{\rho}\NN P
-\frac{n}{\rho}m^A m^B\Bigl( \frac{\FF^{A}(\rr)}
{m^A}- \frac{\FF^{B}(\rr)}{m^B}\Bigl)\Bigl\}, 
\label{adefdiffforce2}
\ee
where  we have introduced the mutual diffusion coefficient $D^{AB}$ (see ref. \cite{Ferziger})
through the linear relation between the velocity and the difference between the chemical potential gradients,
the factor $(k_B T)^{-1}$ having been introduced in the definition  for dimensionality reasons. 
By comparing eqs. \eqref{wapprox} and \eqref{adefdiffforce2} we find
\be
D^{AB} 
=
\frac{k_B T }{\gamma}\frac{\rho}{n} \frac{1 }{m^A m^B },
\label{mutualdiffusion}
\ee
which in the case of equal masses takes the simpler form $D^{AB} 
=\frac{k_B T }{\gamma} \frac{1 }{m}$.
Eq.  (\ref{dab})  relates
a response quantity, the friction coefficient $\gamma$ to a fluctuation quantity,
$D^{AB}$, the mutual diffusion coefficient, according to the  Einstein relation.

Relation \eqref{adefdiffforce2} expresses the fact that the diffusion velocity is opposed to the
gradient of the concentration field (proportional to the first term within the parenthesis)
and that heavier molecules tend  to move towards regions of higher pressure.
The last term in eq. \eqref{adefdiffforce2}  corresponds to the so-called forced diffusion.
We have neglected the Soret effect, that is, the coupling with the temperature gradient, being consistent
with our isothermal treatment. The appropriate extension of the present
theory to thermal systems was
proposed in ref. \cite{molphys2011}.
 Using eq. \eqref{gammasymmetrized1}, for $\gamma$, the mutual diffusion coefficient can be written explicitly as:
\be
D^{AB} 
=\frac{3}{8 n} \frac{  (k_B T)^{1/2} }{(2 \pi \mu^{AB} )^{1/2}(\sigma_{AB})^2 
g^{bulk}_{AB}},
\label{dab}
\ee
an expression identical to that
derived from the Enskog analysis \cite{Chapman}.
Moreover,
assuming that the mass density variations  are negligible and using the relation between the chemical potential difference $ \mu_D$
and the Gibbs free energy per unit mass $G(P,T,n,c)$ \cite{Landau} ,
$$
\frac{\partial G}{\partial c}|_{P,T.n}=\mu_D,
$$
we can write the advection-diffusion equation for the mass concentration 
in the suggestive form
\be
\frac{\partial}{\partial t} c(\rr,t)+\uu \cdot \nabla c(\rr,t) =\frac{1}{\gamma}\nabla
\Bigl [(c(\rr,t)(1-c(\rr,t)) \nabla \frac{\delta G[c]}{\delta c(\rr,t)}  \Bigl]
\label{concdiff0}
\ee
which
 bears a close resemblance with the typical DDFT  equation,
with the Gibbs potential per unit mass $G$ replacing the Helmholtz free energy per unit volume. 
In the case of a binary ideal gas mixture ($\gab^{bulk}=1$) with equal masses,
we recover the standard advection-diffusion equation with a constant diffusion coefficient:
\be
\frac{\partial}{\partial t} c(\rr,t)+\uu \cdot \nabla c(\rr,t) =D^{AB}\nabla^2 c(\rr,t)
\label{concdiff}
\ee
Notice that the present diffusion coefficient $D^{AB}$ corresponds to the Enskog  and not to the Stokes-Einstein expression, since
the underlying dynamics is purely markovian \cite{hansen,molphys2011}.

%%%%%%%%%%%%%%%%%%%%
%%%%%%%%%%%%%%%%%%%%%%

Before concluding this section, we recall that,
 within
the random phase approximation, the presence of attractive tails in the pair potentials 
does not produce any change on the 
coefficients of viscosity and thermal conductivity with respect to their
values in the hard-sphere system. This result is an artifact of the RPA method and  
is worse than the corresponding result obtained via the  Enskog method \cite{Chapman}.
However,  the value of the diffusion coefficient does depend on the potential tails as pointed out in ref. \cite{Stell}.
 In fact, the diffusion current is proportional to  the gradient of the
chemical potential difference $\mu_D$  of eq. \eqref{mudif}.

\begin{figure}[htb]
\includegraphics[clip=true,width=8.0cm, keepaspectratio,angle=0]{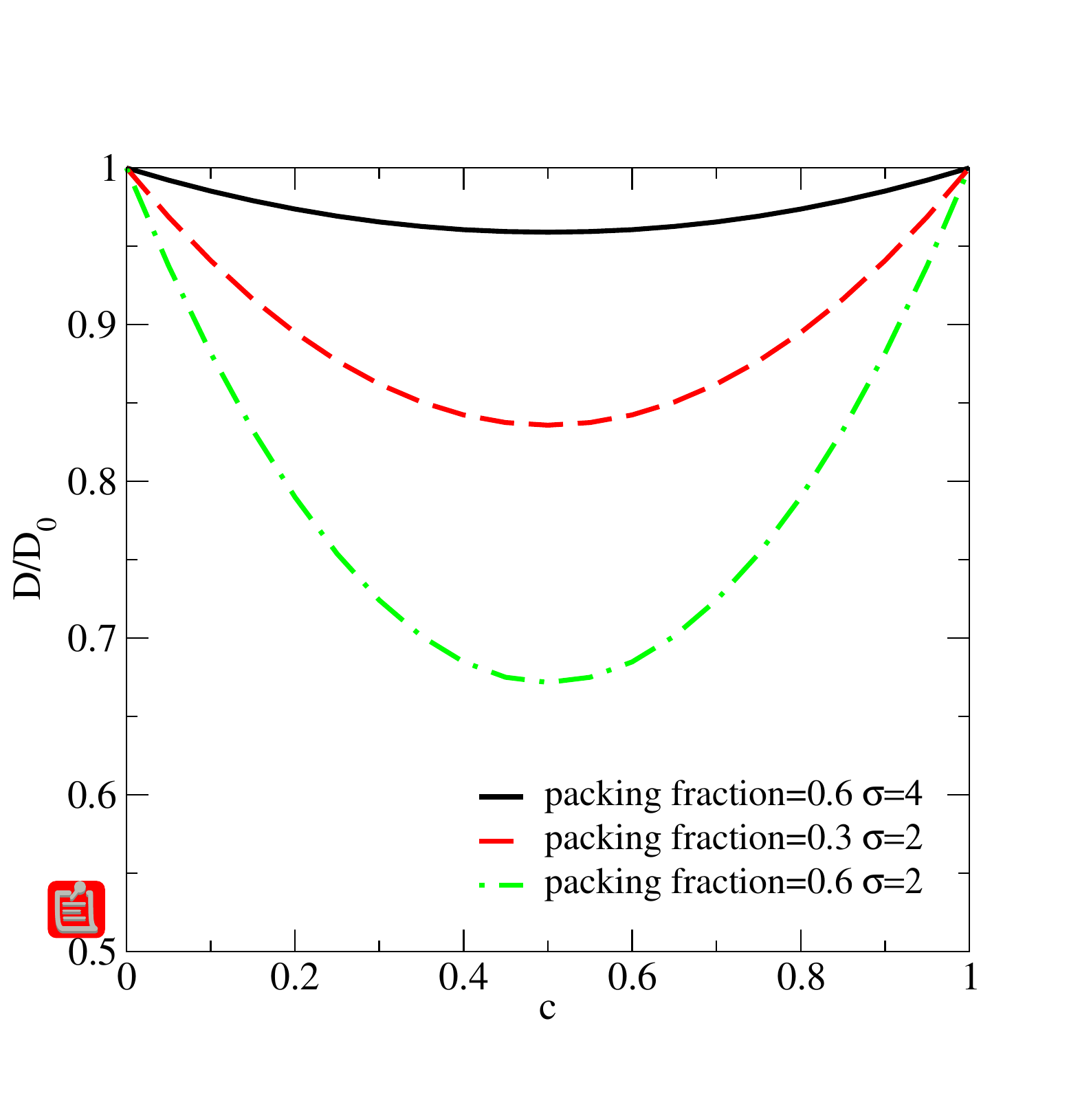}
\caption{Ratio between the mutual diffusion coefficient with attractive tails and 
the coefficient of a system without attractive tails as a function of
concentration.  The mixture consists of  equisized spheres with attractive potentials 
whose strength is chosen according to the Lorentz-Berthelot mixing rule
\eqref{berthelot}. The hard sphere radii are 
$\sigma_{AA}=\sigma_{BB}=2$ in one case and $\sigma_{AA}=\sigma_{BB}=4$
and the packing fraction
$\xi_3=\frac{\pi}{6}(n^A\sigma_{AA}^3+n^B\sigma_{BB}^3)$    is kept fixed at values $0.6$ and $0.3$, 
while varying concentration.
The effect of the potential  tails is the largest for equal concentrations.}
\label{tre}
\end{figure}

In Fig. \ref{tre}, we display the behavior of   the mutual diffusion coefficient  for 
a mixture of equisized hard-spheres
as a function of the bulk concentration for three different  values of the bulk packing fraction. 
The relative strength
of the attractive tails  was fixed empirically
according to the geometric mean Lorentz-Berthelot mixing rule \cite{berthelot}:
\be
w_{AB}=\sqrt{w_{AA} w_{BB}}
\label{berthelot}
\ee
and set $w_{AA}=5 k_B T$ and $w_{BB}= w_{AA}/2$.

We observe that
the presence of attractive  interactions
tends to reduce the value of  $D^{AB}$.
The largest  deviation from the unperturbed value  occurs at 
concentration $c=1/2$ and  decreases at fixed packing fraction as the diameter increases.

%%%%%%%%%%%%%%%%%%%%%%%%%%%%%%%%%%

\subsection{Hydrodynamic analysis}
\label{hydrodynamicanalysis}
We now turn our attention to the case where interspecies diffusion is coupled to acoustic and shear modes.
The following treatment will be based on linearized equations and has the purpose of connecting the
macroscopic hydrodynamic properties, such as the dispersion relations of the propagating modes and their damping, to the
microscopic parameters of the underlying model.
After linearizing eqs. \eqref{densityconservation},\eqref{continuityconcentration}, \eqref{globalmomentumcont}   and \eqref{diffmoment} around the state
$(\rho_0,c_0,\uu=0,{\bf w}=0)$
we find the following set of equations
\bea
&&\partial_{t}\delta \rho(\rr,t)+\rho_0\nabla\cdot  \uu(\rr,t)=0
\label{cc1}
\\
&&\partial_{t} \uu(\rr,t) +\frac{1}{\rho_0}\nabla P(\rr,t)
-\frac{1}{\rho_0}\Bigl(\eta \nabla^2\uu(\rr,t)+(\frac{1}{3}\eta+\eta_b)\nabla(\nabla\cdot\uu(\rr,t))\Bigl)=0
\label{cc2}
\\
&&\frac{\partial}{ \partial t}{\bf w}(\rr,t)+{\bf \nabla}\mu_D(\rr,t)+\gamma
{\bf w}(\rr,t)=0
\label{cc3}
\\
&&\partial_{t}\delta c(\rr,t)+c_0(1-c_0)\nabla\cdot  {\bf w}(\rr,t)=0.
\label{cc4}
\eea 
We now insert the trial solutions, with $\delta\rho,\delta c_0, \uu_0 ,{\bf w_0}$ constants, 
%%   DEVIATIONS
\bea
\delta \rho(\rr,t)&=&\delta \rho_0 e^{\zeta t+i {\bf q \cdot r}}\\
\uu(\rr,t)&=& \uu_0 e^{\zeta t+i {\bf q \cdot r}}\\
\delta c(\rr,t)&=&\delta c_0 e^{\zeta t+i {\bf q \cdot r}}\\
{\bf w}(\rr,t)&=&{\bf w}_0 e^{\zeta t+i {\bf q \cdot r}}\\
\eea
and separate the components of the velocities $\uu$ and ${\bf w}$ into their longitudinal
and transverse parts ($\nabla\times \uu=0$ and $\nabla\cdot \uu=0$, respectively, and similarly for $\ww$.)
Choosing ${\bf q}$ along the z-axis, we rewrite
\bea
&&\zeta\delta \rho_0+i q \rho_0 u^z_0=0
\label{dd1}
\\
&&\zeta u^z_0 +iq\frac{1}{\rho_0} \bigl(\frac{\partial P}{\partial \rho}\bigl)_c \delta \rho_0+
iq\frac{1}{\rho_0} \bigl(\frac{\partial P}{\partial c}\bigl)_{\rho} \delta c_0+q^2\frac{1}{\rho_0}\Bigl(\frac{4}{3}\eta+\eta_b\Bigl)u^z_0=0
\label{dd2}
\\
&&\zeta u^{x(y)}_0 +q^2\frac{\eta}{\rho_0}u^{x(y)}_0=0
\label{dd3}
\\
&&\zeta w^z_0 +iq \bigl(\frac{\partial \mu_D}{\partial \rho}\bigl)_c \delta \rho_0+
iq\bigl(\frac{\partial \mu_D}{\partial c}\bigl)_{\rho} \delta c_0+\gamma w^z_0=0
\label{dd4}
\\
&&\zeta w^{x(y)}_0 +\gamma w^{x(y)}_0=0
\label{dd5}
\\
&&\zeta \delta c_0 +i q c_0(1-c_0)  w^z_0=0,
\label{dd6}
\eea
where the upper indexes indicate Cartesian components of the vectors.
We define the kinematic longitudinal  viscosity $\nu_l=(4 \eta/3+\eta_b)/\rho_0$ and the kinematic 
shear viscosity $\nu=\eta/\rho_0$.
Since the model is isothermal
there is no coupling to the heat modes and
the transverse velocities are completely decoupled from the remaining variables.
As a consequence, the
two shear modes describing standard diffusion of transverse momentum, can be represented as
 \be
u^{x(y)}(\rr,t)= u^{x(y)}_0 e^{-\nu q^2 t+i {\bf q \cdot r}}
\ee
Similarly, the transverse component of the field $\ww$
decays exponentially fast due to the presence of internal friction 
\be
w^{x(y)}(\rr,t)=w^{x(y)}_0 e^{-\gamma t+i {\bf q \cdot r}}.
\ee
The remaining four longitudinal modes are mutually coupled and one has to consider the roots of the determinant 
%%%%%%% DETERMINANT
\begin{large}
$$
\left| \begin{array}{cccc} \zeta & iq\rho_0  & 0&0\\ 
\frac{iq}{\rho_0}(\frac{\partial P}{\partial \rho})_c & \zeta+\nu_l q^2 & 0 & \frac{iq}{\rho_0}(\frac{\partial P}{\partial c})_\rho \\
iq(\frac{\partial \mu_D}{\partial \rho})_c  & 0 & \zeta+\gamma & iq(\frac{\partial \mu_D}{\partial c})_\rho \\
0& 0 & iq c_0(1-c_0)&\zeta\end{array} \right|
$$
\end{large}
% end Determinant
For the hydrodynamic analysis, it is sufficient to compute the roots 
of the associated  fourth order   secular equation  to order $q^2$,
so to obtain the following roots:
\be
\zeta_{acoustic}= \pm i c_s q-\Gamma q^2\ee
with a sound velocity given by
\be
c_s=  \sqrt{(\frac{\partial P}{\partial \rho})_c}
\ee
and where
\be
\Gamma=-\frac{1}{2}\Bigl(\nu_l+(\frac{\partial  \mu_D}{\partial \rho})_c (\frac{\partial P}{\partial c})_\rho/
(\frac{\partial P}{\partial \rho})_c \Bigl)
\label{damping}
\ee
The last term in eq. \eqref{damping} represents the damping of sound waves by interdiffusion of the two species.
Finally the species diffusion is associated with the eigenvalue
\be
\zeta_{diffusive}=-D_0 q^2
\ee
with
\be 
D_0\equiv\frac{1}{\gamma}c_0(1-c_0) \frac{\partial \mu_D}{\partial c}
\ee
which should be compared with the r.h.s of  eq.\eqref{concdiff0}.
  %%%%%%%%%%%%%%%%%%%%%
%%%%%%%%%%%%%%%%%%%%%%%%
\section{Numerical validation }
\label{Numerical}
 In this section we compare some of  the theoretical predictions   with the numerical results obtained 
by applying the Lattice Boltzmann  numerical solution of the coupled kinetic
equations \eqref{evolution}. The discretized form of these equation has been presented  in detail in appendix B
of ref. \cite{JCP2011} and will not be repeated here for the sake of brevity.
 \subsection{Molecular diffusion}
 \label{sub:standarddiffusion} 
 We first determine the diffusion coefficient of an hard-sphere mixture at various packing fractions and compositions
and then consider the effective diffusion coefficient for a system subject to a special type of shear flow.
In the case of small perturbations around the equilibrium state, it is possible to obtain an analytical estimate of the
so-called Taylor dispersion \cite{taylor}.

\begin{figure}[htb]
\includegraphics[clip=true,width=8.0cm, keepaspectratio,angle=0]{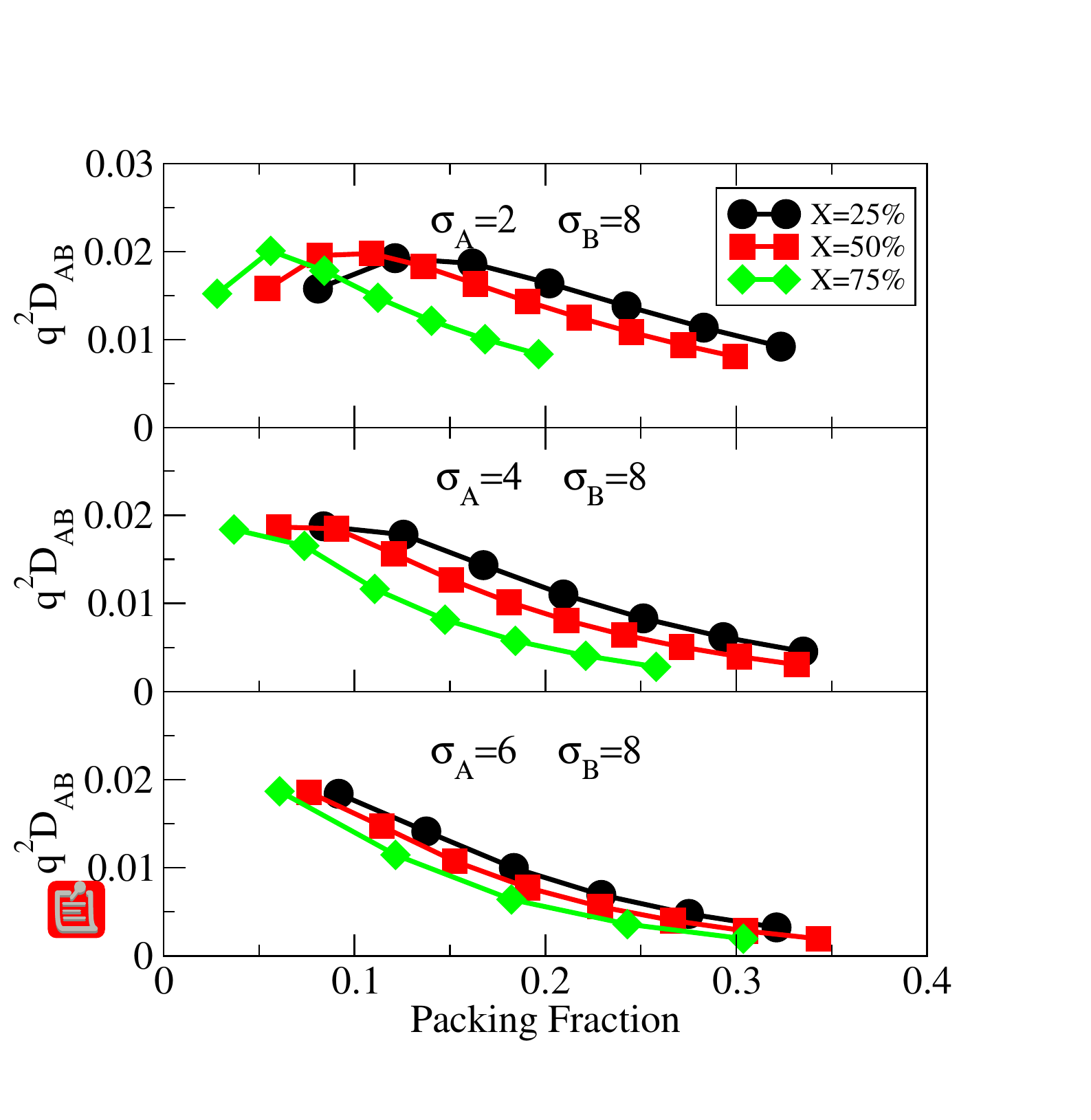}
\caption{Numerical test of the diffusion process in bulk conditions. The vertical axis represents a measure of  the mutual diffusion coefficient
obtained from LBM simulations (all data expressed in LBM timestep units). We  monitored the evolution towards equilibrium of a sinusoidal concentration
fluctuation (see eq. \eqref{perturba}) of wave-vector $q_z$ and  extracted 
the characteristic  decay  time, $1/\tau(q_z)=D^{AB} q_z^2$.
 The plots report the inverse decay time versus packing for various values of the composition and diameter ratio
 and for a fixed value of $q_z=40$.}
\label{bmutual}
\end{figure}

Let us first consider the relaxation of an initial concentration gradient in a system with $m^A=m^B$, $\uu=0$ and whose 
global density is uniform.
In the initial state  the composition varies along the $z$ direction as a sinusoidal wave of small amplitude, $\Delta$,
and given by the two distribution functions
\bea
f^A(\rr,\vv,t=0) &=& (n_0+\Delta \sin(q_z z) )e^{-m \vv^2/(2 k_B T)} \nonumber \\
f^B(\rr,\vv,t=0) &=& (n_0-\Delta \sin(q_z z) )e^{-m \vv^2/(2 k_B T)} .
\label{perturba}
\eea
The diffusion constant is computed by monitoring the decay of a particular peak of $n_A$, which
according to the theory, decreases exponentially  with an inverse characteristic time $1/\tau(q_z)=D^{AB} q_z^2$.
The extracted value of $D^{AB}$ as a function of the packing fraction for several values of the bulk composition
and different diameter ratios is reported in  Fig. \ref{bmutual}.

We observe  that the mutual diffusion coefficient, $D^{AB}$, increases as the concentration of large spheres increases at fixed value of the
packing fraction, according to the theoretical prediction eq.(\ref{dab}).
On the other hand, at fixed concentration and high packing fractions, the diffusion constant decreases as a function of the packing fraction.

However, in the low density region we find the unexpected result that the diffusion constant increases with the
packing fraction. This regime correlates with the fact that
the decay  of the perturbation \eqref{perturba} 
does not decay diffusively, but displays an oscillatory damped behavior.

This phenomenon occurs when the wavevector
of the initial fluctuation is larger than a critical value $q_c= \sqrt{\gamma/4 D}$.
This apparent deviation from the standard diffusive behavior
is the result of probing the system at small scales where standard hydrodynamics does not apply.
However, since the phenomenon occurs only at finite wavelength below
a certain threshold it is not in contradiction with
the hydrodynamic picture presented above.
The diffusion equation obtained in section \ref{homogeneousdiffusion} holds in the hydrodynamic regime 
when, as a result of many collisions, the fluid has reached local equilibrium.
In terms of wave-vector and frequency one requires $q\lambda_{mf}<<1$ and
$\omega \tau<<1$ , where $\lambda_{mf}$  is the mean free path and $\tau$ the
mean collision time. At densities typical of a liquid  the mean free path is of the
order of magnitude of the molecular size, while in a very diluted gas $\lambda_{mf}$ 
becomes large so that the range of validity of hydrodynamic formulae shrinks.

The following simple analysis shows the origin of the non monotonic decay.
We first decouple  the ``acoustic'' modes in the 
hydrodynamic matrix, by neglecting the derivative of the
pressure with respect to concentration, and consider the simplified equation 
( by neglecting $(\frac{\partial P}{\partial c})_\rho\simeq 0$):
\be
\zeta^2+\gamma\zeta+\Delta q^2=0
\ee
with $\Delta=c_0(1-c_0) (\frac{\partial \mu_D}{\partial c})_\rho$.
The following decay frequencies:
\be
 \zeta_{\pm}=-\frac{\gamma}{2}\pm\sqrt{\frac{\gamma^2}{4}-\Delta q^2}
 \ee
display oscillatory-damped  behavior for concentration fluctuations
of  wave-vectors $q>q_c$, with $q_c=\sqrt{\frac{\gamma^2}{4 \Delta}}$. 
 At low density we can obtain an analytic expression for such a crossover, since
$ \Delta  \simeq \frac{k_B T}{m}$ 
 and
 \be
\gamma
=\frac{8}{3}\frac{\sqrt{ k_B T }}{\sqrt {2\pi m}} \frac{1} \lambda_{mf} ,
\label{gammasymmetrized}
\ee
where the mean free path is
$\lambda_{mf}=\frac{1}{\sqrt 2 \pi g(\sigma)  \sigma^2 n}$. 
In terms of the wavelength $L_c=2\pi/q_c$
 the transition from the diffusive to the oscillatory damped
behavior occurs when the Knudsen number, expressing  the ratio of the two
characteristic lengths of the problem, is
   \be
Kn =\frac{  \lambda_{mf} }{L_c}=  \frac{4}{3}\frac{1}{(2\pi )^{3/2}} \simeq 0.08 .
   \label{critical}
     \ee 
   In other words, if the wavelength of the fluctuation is
of the order of the mean free path, collisions
are not frequent enough to restore local equilibrium, which is the mechanism
determining molecular diffusion.

Such an oscillatory
decay of diffusive modes should be contrasted
with the behavior associated to a simple BGK
collisional kernel, which does not have such oscillations \cite{Asinari,luoasinari,guoasinari}. In fact, in the latter
the friction constant, $\gamma$, is not determined self-consistently
but enters as a free parameter and is usually assumed to be 
a density independent quantity.

\subsection{Taylor dispersion in a periodically modulated flow}

In this subsection, we discuss a problem where one observes 
the interplay of  a macroscopic  and microscopic mechanisms.
This occurs, for instance, when an inhomogeneous concentration field is subjected to a non-uniform 
macroscopic velocity flow. As discovered by Taylor such a situation
determines  an enhancement of the molecular diffusion in the direction of the flow, known as 
Taylor dispersion \cite{taylor}.

The theoretical calculation, sketchly reported hereafter  for the sake of completeness, is based on a multiscale 
perturbation analysis. We refer  to the work of \cite{Yanna} for mathematical details. 

We consider a  box of length $L_x$ and 
cross-section $L_y\times L_z$ and  a fluid velocity $u^x(y)$ periodically modulated along the $y$ 
direction:
\be
\uu (\rr)=\left(-U \cos\left(\frac{2\pi y}{L_y}+\pi\right),0,0\right)
\label{ushape}
\ee
%and periodic along the $y$ direction within the interval  $0<y<L_y$.
The boundary conditions are such that at the extremes of the box $u^x=-U$,
and at the center of the box $u^x=U$.

 In the diffusive regime the concentration obeys the three dimensional advection-diffusion equation \eqref{concdiff}
with $\uu$ given by \eqref{ushape}.
However, the description can be contracted, using multiscale techniques,  and instead of studying the evolution of the full 
concentration field
one can focus attention on the sectionally averaged concentration, $C(x,t)$ given by:
\be
C(x,t)=\frac{1}{L_y}\int_{0}^{L_y} d y c(x,y,t)
\ee
in the presence of a laterally averaged velocity
\be
U_0=\frac{1}{L_y}\int_{0}^{L_y} d y u^x(y).
\ee
The theory \cite{Yanna}  shows that the salient information about the diffusive process is given by   
the following simpler  one-dimensional advection-diffusion equation:
\be
\frac{\partial}{\partial t} C(x,t) +U_0 \frac{\partial}{\partial x} C(x,t)
=D_{eff}  \frac{\partial^2}{\partial x^2}C(x,t)
\label{c99}
\ee
where the new  coefficient $D_{eff}$ is due to the  renormalization of the standard molecular diffusion 
induced by the macroscopic velocity field $u^{x}(y)$. Its value is given by the formula:
\be
D_{eff}=D^{AB}( 1+\frac{U^2 L_y^2}{8\pi^2 (D^{AB})^2}) 
=D^{AB} (1+\frac{Pe^2}{2\pi^2}) 
\label{taylordiff}
\ee
where the Peclet number, $Pe=U L_y/2 D^{AB}$, is the ratio between the rate of advection and the rate of molecular diffusion.

\begin{figure}[htb]
\includegraphics[clip=true,width=8.0cm, keepaspectratio,angle=0]  {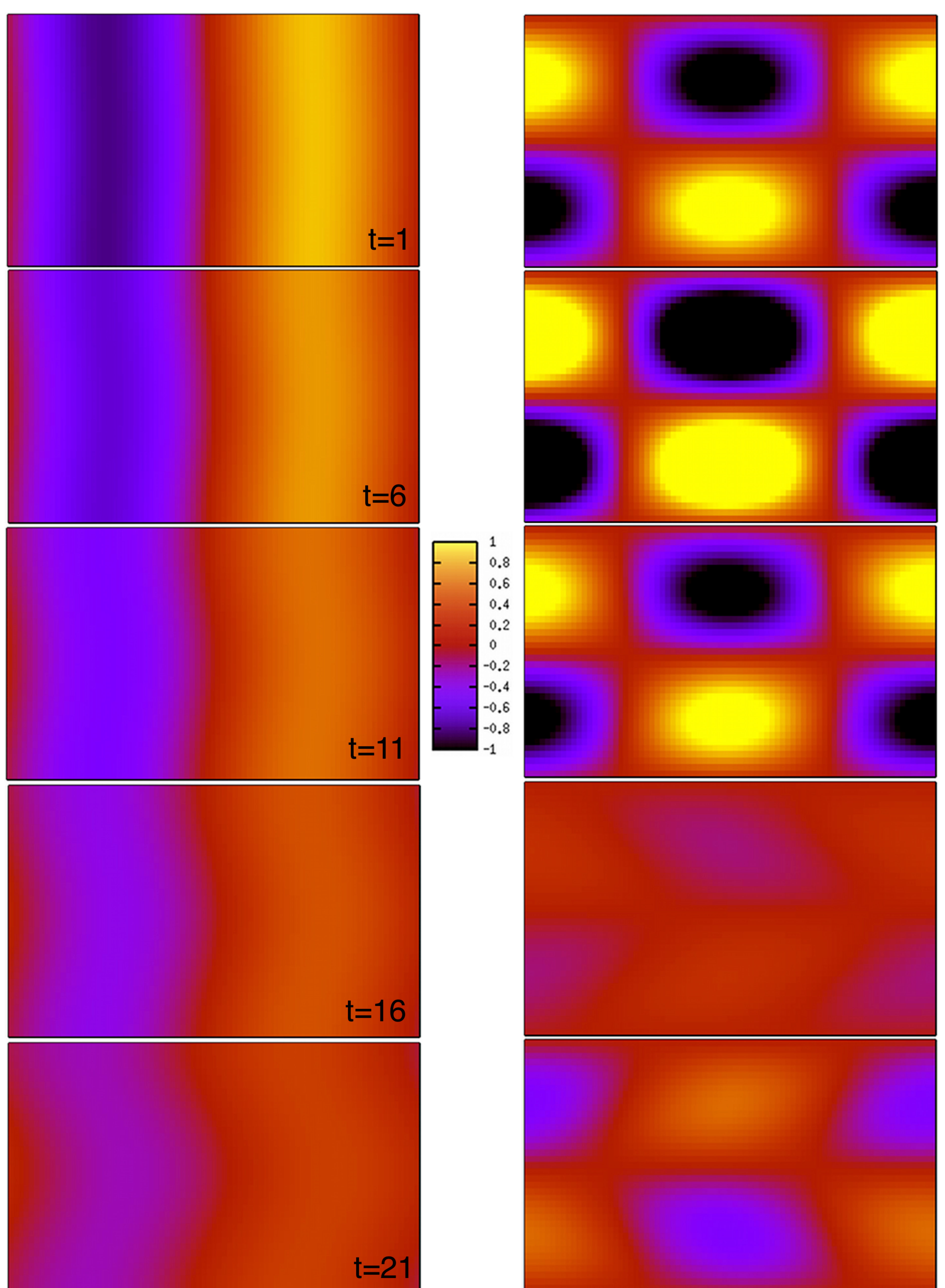}
\caption{Time evolution of the density (left column) and current (right column) in Taylor dispersion. 
The initial concentration modulation is along the $x$ direction whereas the external field varies
along the $y$ direction according a cosine law. 
In the left column we report the evolution of the density of the large species every $5$ LBM timesteps.
In the right column we report the evolution of the associated current in the $y$ direction. 
Data correspond to $\sigma_{AA}=8$, $\sigma_{BB}=4$, $c_0=0.5$, $Pe=1$, average packing $\xi_3=0.211$, 
and for a simulation box of $80\times40\times40$. The color scale refers to both the
density and current plots. Both reported data are normalized according to the initial values 
of the respective fields.}
\label{pool1}
\end{figure}

\begin{figure}[htb]
\includegraphics[clip=true,width=8.0cm, keepaspectratio,angle=0]  {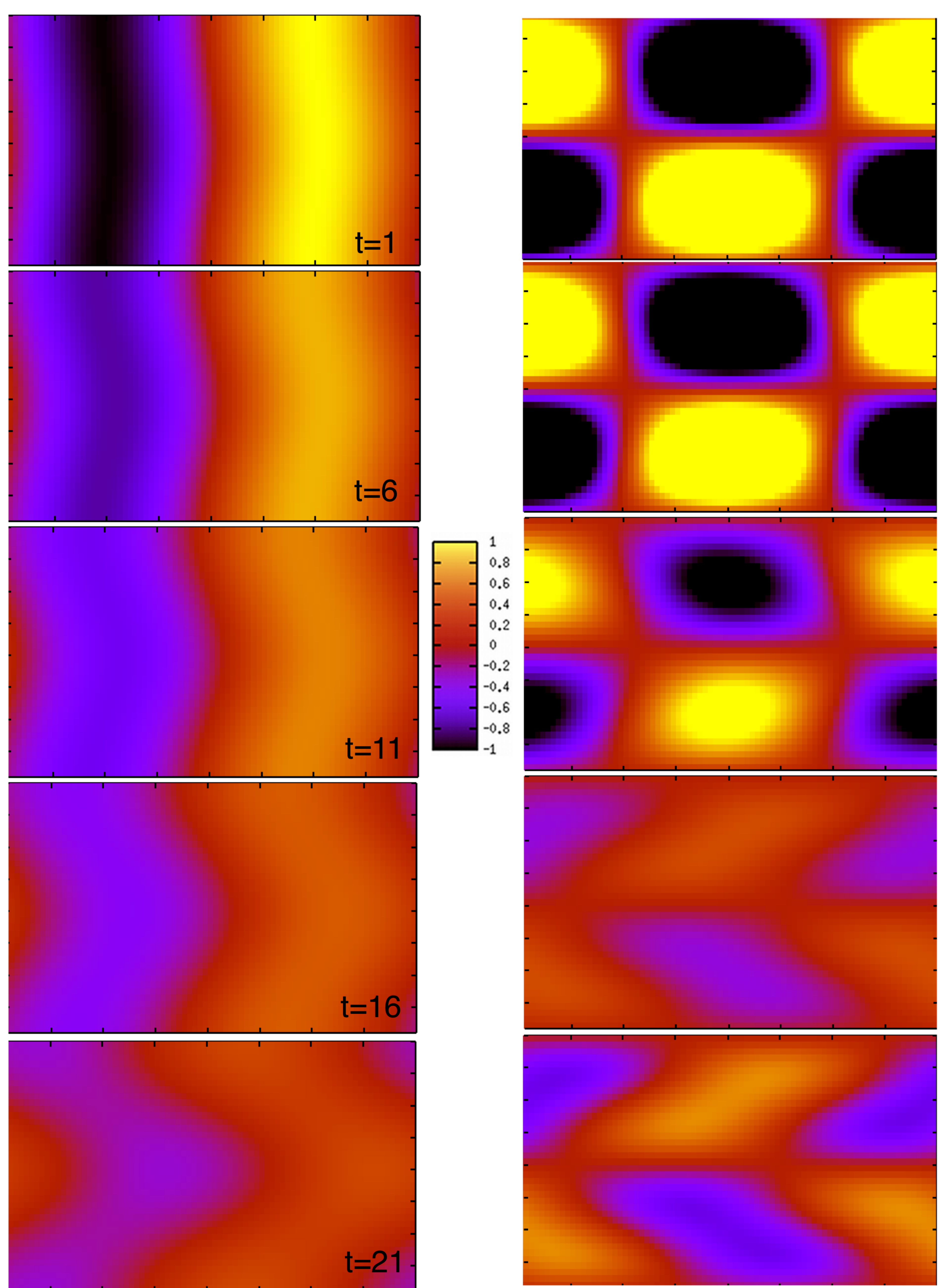}
\caption{Time evolution of the density  (left) and current  (right) in Taylor dispersion. 
The fluid parameters are the same as in Fig.~\ref{pool1}, but for Peclet number $Pe=5$.}
\label{pool2}
\end{figure}

So far goes the theory.  The above scenario can be checked with a numerical calculation  similar to that of sub-section \ref{sub:standarddiffusion} appropriately modified in order to account for the presence of the field $\uu(\rr)$.
We assumed
an initial  concentration inhomogeneity
along the $x$ direction
under the form of
two initial density  fields:
$$
n^A(x)=n^A_0+\Delta \sin(k_x x)
$$
$$
n^B(x)=n^B_0-\Delta \sin(k_x x).
$$
 and verified that  the homogeneous state is recovered exponentially with  a characteristic time 
$1/\tau(q)=D^{AB}_{eff} q^2$,
that depends on the strength of the applied velocity field  and its  wavelength as predicted by \eqref{taylordiff} .
In Figs. \ref{pool1} and \ref{pool2} we display the different stages of the evolution of the concentration field in the left columns and of the velocity
field in the right columns, obtained from our LBM code, for two different values of the strength of the imposed velocity field.
One can see that the concentration gradient tends to decrease as the time increases and so does the concentration current. 
In a matter of $\sim 30$ LBM timesteps the concentration gradient is barely visible and the currents have faded away. During
its evolution, the density field initially distorts in a quasi-parabolic shape and subsequently in a v-shaped form, being
more pronounced at high Peclet. The current displays non-trivial patterns alternating in time and position as time proceeds.

\begin{figure}[htb]
\includegraphics[clip=true,width=8.0cm, keepaspectratio,angle=0]{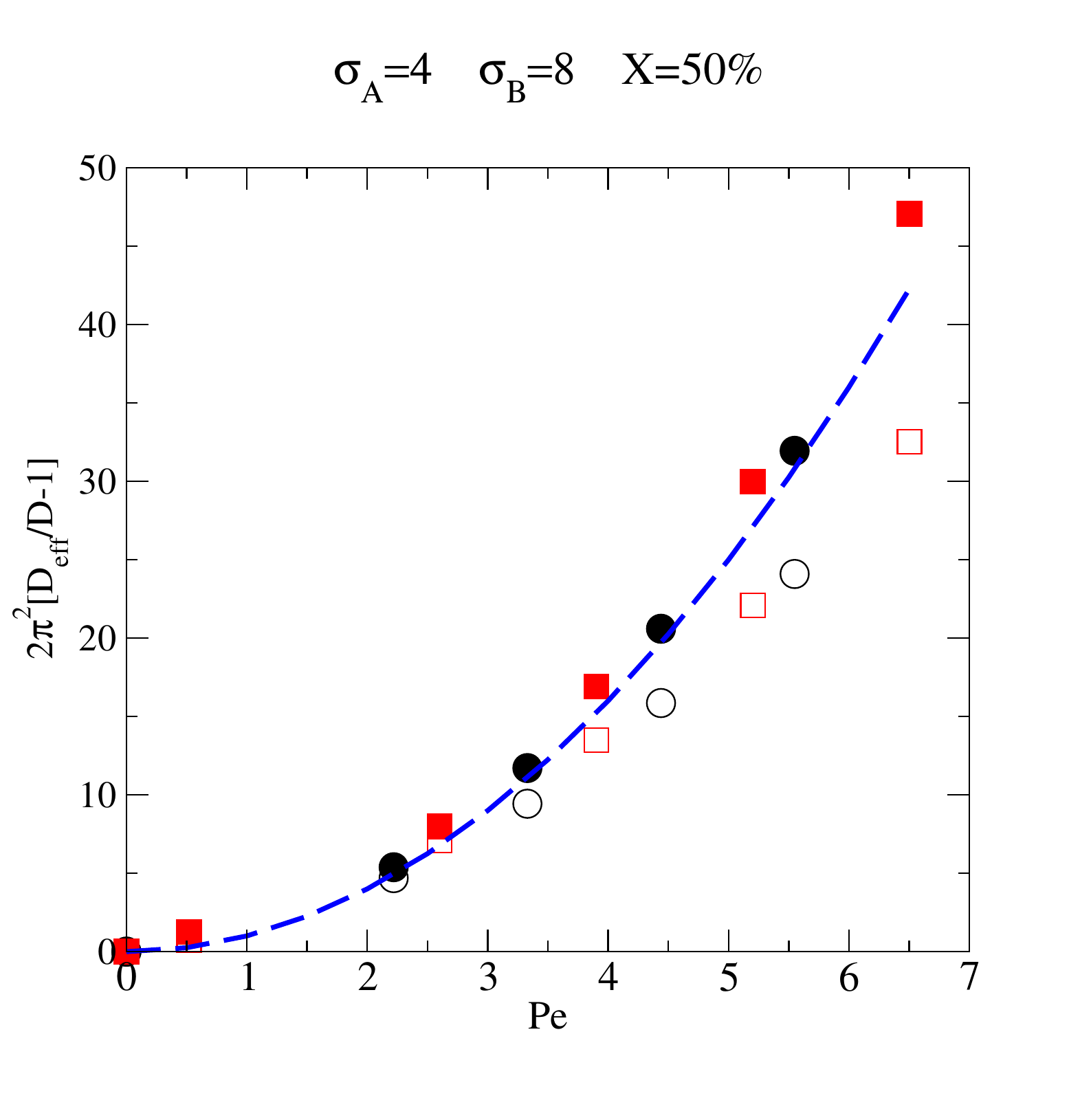}
\caption{Mutual diffusion coefficients in Taylor dispersion
obtained for packing fraction of 0.211 (circles) and 0.332 (squares respectively). 
Filled symbols correspond to a box of length $L_x=80$, while open symbols to a box of length $L_x=40$.
The dashed line corresponds to eq. (\ref{taylordiff}).  }
\label{mutual}
\end{figure}

The numerical results obtained from our simulation are checked against theoretical predictions, 
derived under the  assumption that
the concentration field is assimilable to a passive scalar. 
Fig. \ref{mutual} displays the relaxation time for a concentration inhomogeneity for various values
of the Peclet number, and for two values of the packing fraction.
The effective diffusion increases quadratically as a function of the Peclet number as predicted by the theory.

%%%%%%%%%%%%%%%%%%%%%%%%
%%%%%%%%%%%%%%%%%%%%%%%%%
\section{Conclusions}
\label{Conclusion}

Using a microscopic approach based on the multicomponent Boltzmann- Enskog  equation
and a self-consistent treatment of the interactions
we have studied the diffusional properties of a mixture of hard-spheres.
In order to obtain a working scheme we have employed  a series of
hypothesis and approximations. First, we have assumed that the complex many body problem can be represented
by means of a modified Boltzmann-Enskog equation, the RET, where only configurational two-particle correlations are
accounted for. Since the RET requires a reasonable effort only in the case of hard-sphere interactions, the attractive potential
tails have been treated within the RPA, an approximation which fails to accurately reproduce the transport coefficients.
A second important simplification adopted is the method of Santos et al. \cite{Brey}, where
the slowly varying, hydrodynamic fields and the fast non-hydrodynamic ones are decoupled at kinetic level, and
the latter are treated in a simplified way. 
Third, we only considered isothermal situations for the sake of simplicity. The extension to 
non-isothermal systems will be the subject of future work.
Finally, we have discretized 
the resulting transport equations on a lattice and employed the Lattice Boltzmann method to obtain numerical solutions.

The present method represents a valid alternative 
to popular mesoscopic techniques, such as the 
pseudo-potential-Lattice-Boltzmann (LB) method or free-energy based models (see for instance, ref. \cite{LBgeneral} and
references therein),
that retain the functional form of
the equilibrium free energy, but sacrifice  the possibility
of determining
the transport from the  microscopic
pair potentials  through controlled approximations.
In contrast, our approach leads in a quite natural fashion 
to the determination of thermodynamic forces compatible
with the free energy methods, but in addition
determines self-consistently the non equilibrium forces necessary
to guarantee the correct hydrodynamic behavior.

We have obtained a derivation of the advection-diffusion equation for the concentration and the
self-consistent determination of the diffusion coefficient, which in the homogeneous case reduces to the
Chapman-Enskog value. The study of the long wavelength and low frequency properties of the  model has been performed 
and agrees with the results obtained by standard hydrodynamic analysis \cite{hansen,Boon} of mixtures.

A second merit of the present formulation is to lend itself to
numerical solution via the Lattice Boltzmann method.
Our computational  approach takes into account the dynamics of flowing 
liquids on space-time scales of hydrodynamic interest. These scales are
out of reach for Molecular Dynamics, which in principle
describes {\em ab-initio} the system, 
since the probabilistic nature of the singlet distribution function 
does not require averaging the data as in particle-based methods. In addition, the proposed
method can cope very naturally with the critical situations of low concentrations of one species.

By simple numerical experiments, we have verified that the  present version of the LBM allows 
to extract the value of the diffusion coefficient 
from the decay of small periodic concentration fluctuations. Moreover,  we considered 
cases where the Taylor dispersion mechanism provides an enhancement of diffusion, 
thus further showing that the present numerical scheme 
is capable of handling molecular mechanisms together with driving forces acting on much 
larger scales.

We plan future applications of the present approach to the study of non-uniform substrates, multiphase flows and transport in narrow channels.

%%%%%%%%%%%%%%%%%%%%%%%%%%%%%%%%%%%%%%%%%
%%%%%%%%%%%%%%%%%%%%%%%%%%%%%%%%%%%%%%%%%
%%%%%%%%%%%%%%%%%%%%%%%%%%%%%%%%%%%%%%%%%

%%%%%%%%%%%%%%%%%%%%%%%%%%%
%%%%%%%%%%%%%%%%%%%%%%%%% 

%%%%%%%%%%%%%%%%
%%%%%%%%%%%%         Fig 1
%--------------------------------------------------------------------------------
%----------------------------- Fig.2 --------------------------------------
%--------------------------------------------------------------------------------%%%%%%%%%%  FIG 3

%%%% FIG 4

%%%%%%%%%%%%%%%%
%%%%%%%%%%%%         Fig 5

%%%%%%%%%%%%%%%%%%%%%%%


\begin{thebibliography}{99}

\bibitem{sparreboom}
W. Sparreboom, A. van den Berg and J. C. T. Eijkel ,
Nature Nanotechnology {\bf 4}, 713, (2009).
\bibitem{bocquet}
L.Bocquet and E.Charlaix, Chem. Soc. Rev. (2009).
\bibitem{schoch}
R.B. Schoch, J.Han and P. Renaud,
Rev.Mod.Phys. {\bf  80}, 839 (2008).
 \bibitem{Bruus}
H. Bruus, {\it Theoretical Microfluidics}, Oxford University Pres., New York, 2008 .


\bibitem{rauscher}
M. Rauscher and S. Dietrich,
Annual Review of Materials Research
 {\bf 38}, 143 (2008).

\bibitem{ArcherEvans}
A.J.~Archer and R.~Evans, J.~Chem.~Phys.~{\bf 121}, 4246 (2004).

\bibitem{Tarazona}
U. Marini Bettolo Marconi and P.Tarazona, 
J. Chem. Phys. {\bf 110}, 8032 (1999) and J.Phys.:  Condens. Matter 
{\bf 12}, 413 (2000).

\bibitem{Lowen}
H. Lowen, J.Phys. Condens.Matt {\bf 14}, 11897 (2002).

\bibitem{rauscher2}
M. Rauscher, J.Phys.:  Condens. Matter  {\bf 36}, 364109 (2010).

\bibitem{wu}
Jianzhong Wu and Zhidong Li, Annual Review of Physical Chemistry
85, {\bf 58} (2007).

\bibitem{Garcia}
F.J. Alexander, A. L. Garcia and B.J. Alder,
Phys.Rev. Lett. {\bf 74}, 5212 (1995).
\bibitem{Santos}
J.M. Montanero and
A. Santos, Phys.Rev. E {\bf 54}, 438 (1996).


\bibitem{umbmnogalilei}
U. Marini Bettolo Marconi and P. Tarazona,
J. Chem. Phys. {\bf 124},  164901 (2006),
U. Marini Bettolo Marconi,  P.Tarazona and F. Cecconi,
J. Chem. Phys. {\bf 126},  164904 (2006),
U. Marini Bettolo Marconi and S.Melchionna,
J. Chem. Phys. {\bf 126},  184109 (2007).


\bibitem{Espanol}
J.G. Anero and P.Espa\~nol,
Europhys.Lett {\bf 78}, 50005 (2007).

\bibitem{Archer2009}
A. J. Archer
J. Chem. Phys. {\bf 130},  014509 (2009).

\bibitem{Lausanne2010}
U. Marini Bettolo Marconi and S.Melchionna, J.Phys.:  Condens. Matter  {\bf 36}, 364110 (2010).

 
\bibitem{Zhaoli}

Zhaoli Guo and T.S. Zhao, Phys. Rev. E {\bf 68}, 035302 (2003).

\bibitem{Greci}
E.S. Kikkinides, A.G. Yotis, M.E. Kainourgiakis and A.K. Stubos,
Phys. Rev.E, {\bf 78}, 036702 (2008) and Phys. Rev.E, {\bf 82}, 056705 (2010).


\bibitem{LBgeneral}

S. Succi,
{\it The Lattice Boltzmann equation for fluid dynamics and beyond},  
1th edition , Oxford University Press, (2001).
\bibitem{shanchen}
X. Shan and H. Chen, Phys. Rev. E {\bf 49}, 2941 (1994).
\bibitem{Doolen}
S. Chen and G. D. Doolen, Annual Review of Fluid Mechanics~{\bf 30}, 329 (1998). 

\bibitem{Luo}
L.S. Luo and S. Grimaji, Phys. Rev E {\bf 66}, 035301 (2002)
and Phys. Rev E {\bf 67}, 036302 (2003).


\bibitem{Melchionna2008}
S.~Melchionna and U. Marini Bettolo Marconi,
Europhys.Lett {\bf 81}, 34001 (2008).

\bibitem{Melchionna2009}
U. Marini Bettolo Marconi and S.Melchionna,
J. Chem. Phys. {\bf 131},  014105 (2009).

\bibitem{JCP2011}
U. Marini Bettolo Marconi and S.Melchionna,
 J. Chem. Phys {\bf 134}, 064118 (2011).



\bibitem{Davis1}
T.K. Vanderlink and H.T. Davis,
J.Chem.Phys. {\bf 87}, 1793 (1987).
\bibitem{Davis2}
J.J. Magda, M.V. Tirrel and H.T. Davis,
J.Chem.Phys. {\bf 83}, 1888 (1985).

\bibitem{Pozhar}
L.A. Pozhar and K.E. Gubbins, J.Chem.Phys. {\bf 94}, 1367 (1991).

 \bibitem{nicholson}
D. Nicholson and S.K. Bathia,
Molecular simulation {\bf 35}, 109, (2009).

 \bibitem{Rice}
 S.A. Rice and A. Allnatt,
 J. Chem. Phys. {\bf  34},  2144 (1961).
 
 
 \bibitem{Brey}
J. W. Dufty, A. Santos, and J. Brey, Phys. Rev. Lett. ~{\bf 77}, 1270 (1996) and
A.Santos, J.M. Montanero, J.W. Dufty and J.J. Brey, Phys.Rev. E ~{\bf 57}, 1644 (1998).
\bibitem{taylor}
G.I. Taylor,  Proc. Phys. Soc. B 67, 857 (1954).
\bibitem{vanbeijeren}
H. van Beijeren and M.H. Ernst, Physica A, {\bf 68}, 437 (1973),
{\bf 70}, 225 (1973).

\bibitem{Lopezdeharo}  
M. Lopez  de Haro, E.G.D. Cohen and J.M.Kincaid,
J.Chem.Phys. {\bf 78}, 2746 (1983).


 
 \bibitem{BGK}  
P. L. Bhatnagar, E. P. Gross, and M. Krook, Phys. Rev. 
{\bf 94}, 511 (1954).
 
\bibitem{hansen}

J.P. Hansen and I.R. McDonald, {\it Theory of Simple Liquids}
Academic Press, Oxford,  (1990).

\bibitem{Evans1}
R.~Evans, Adv.~Phys.~{\bf 28}, 143 (1979).

\bibitem{Carey}
B.S. Carey and L.E. Scriven. J. Chem. Phys.  {\bf 69}  5040 (1978).
 %%%%%%%%%%%%%%%
 \bibitem{LeeFischer}
 T. Lee and P. F. Fischer, Phys. Rev. E 74, 046709 (2006).
\bibitem{molphys2011}

U. Marini Bettolo Marconi, Mol.Phys. {\bf 109}, 1265 (2011).

%%%%%%%%%%%%%%%%%%%%%%%%%
\bibitem{Chapman}
S. Chapman and T.G. Cowling {\it The mathematical Theory  of Non-uniform Gases}
3rd ed. (Cambridge University Press, Cambridge 1970)

\bibitem{Degroot}
S.R. De Groot and  P. Mazur ,  {\it Non-Equilibrium Thermodynamics}. New 
York, NY: Dover Publications; 1984.  

\bibitem{Ferziger}
J.H. Ferziger, H.G. Kaper, {\it Mathematical Theory of Transport Processes in gases}, Noth-Holland, Amsterdam, 1972.

%%%%%%%%%%%%%%%%%%%%%%%%%%%%%%%%
\bibitem{Pina}
E. Pi\~na, L.S. Colin and P. Goldstein, Physica A {\bf 217},  87 (1995).

\bibitem{Tham}
M. K. Tham and K. E. Gubbins,J. Chem. Phys. {\bf 55} ,268 (1971).

\bibitem{Gross}
E. P. Gross, and M. Krook, Phys. Rev. {\bf 102}, 593 (1956).

\bibitem{Garzo}
V. Garzo, A. Santos and J.J. Brey,
Phys.Fluids A {\bf 1},  380 (1989).

\bibitem{Andries}
P. Andries, K. Aoki and B. Perthame
J. Stat. Phys.~{\bf 106}, 993 (2002).


%BGK models


\bibitem{Landau}
L.D.Landau and E.M. Lifshitz, {\it Fluid Mechanics}, Pergamon Press, London (1963).
%%%%%%%%%%
\bibitem{Stell}
J. Karkheck , E. Martina and G. Stell, Phys.Rev.A 25, 3328, (1982).

\bibitem{berthelot}
M.P. Allen  and  D.J. Tildesley,  {\it Computer Simulation of Liquids} , Clarendon
Press, Oxford (1989).

\bibitem{Asinari}
P. Asinari,
Physics of Fluids, {\bf 17} , 067102 (2005).
\bibitem{luoasinari}
P.Asinari and L.S. Luo, J. Comput. Phys, {\bf 227}, 3378 (2008).
\bibitem{guoasinari}
Z. Guo, P. Asinari and C. Zheng,
Phys.Rev. E {\bf 79}, 026702 (2009).
\bibitem{Yanna}
A. N. Yannacopuolos, G. Rowlands and G.P. King,  J. Phys. Math Gen 31, 377 (1998) .
\bibitem{Boon}
 J.P. Boon and S. Yip, {\it Molecular Hydrodynamics}, Dover, (New York) (1991).


\end{thebibliography}
\end{document}